\documentclass[conference]{IEEEtran}
\IEEEoverridecommandlockouts
\usepackage{cite}
\usepackage{amsmath,amssymb,amsfonts}
\usepackage{algorithmic}
\usepackage{graphicx}
\usepackage{textcomp}
\usepackage{xcolor}
\usepackage[a4paper, total={184mm,239mm}]{geometry}
\def\BibTeX{{\rm B\kern-.05em{\sc i\kern-.025em b}\kern-.08em
    T\kern-.1667em\lower.7ex\hbox{E}\kern-.125emX}}

\usepackage{booktabs} 
\usepackage{gensymb}
\usepackage[english]{babel}
\usepackage{bm}
\usepackage[capitalise]{cleveref}
\usepackage{subcaption}
\usepackage{pifont}
\usepackage{makecell}
\usepackage{multirow}
\usepackage{multicol}

\newcommand{\cmark}{\ding{51}}%
\newcommand{\xmark}{\ding{55}}%

\newcommand{\Vmin}{V_{min}}
\newcommand{\Fmax}{F_{max}}

\newcommand{\loss}{\mathcal{L}}
\newcommand{\lossp}{\loss_p}
\newcommand{\lossq}{\loss_q}
\newcommand{\model}{g}
\newcommand{\modelPoint}{\model_p}
\newcommand{\modelRegion}{\model_r}
\newcommand{\x}{\bm{\mathrm{x}}}
\newcommand{\y}{\mathrm{y}}
\newcommand{\X}{\bm{\mathrm{X}}}
\newcommand{\Y}{\bm{\mathrm{y}}}

\newcommand{\RD}{\mathbb{R}^D}
\newcommand{\R}{\mathbb{R}}

\newcommand{\paramModel}{\bm{\theta}}

\newcommand{\paramLow}{\paramModel_{lo}}
\newcommand{\paramHigh}{\paramModel_{hi}}

\newcommand{\traindataset}{\mathcal{D}}

\newcommand{\Rtwo}{R^2}
\newcommand{\coverage}{\alpha}
\newcommand{\Prob}{\mathbb{P}}
\newcommand{\cilow}{\modelPoint}
\newcommand{\cihigh}{\modelPoint}
\newcommand{\qlow}{q_{lo}}
\newcommand{\qhigh}{q_{hi}}

\DeclareMathOperator*{\argmin}{arg\,min}

\makeatletter
\newcommand{\smallbullet}{} 
\DeclareRobustCommand\smallbullet{%
  \mathord{\mathpalette\smallbullet@{0.7}}%
}
\newcommand{\smallbullet@}[2]{%
  \vcenter{\hbox{\scalebox{#2}{$\m@th#1\bullet$}}}%
}
\makeatother

\begin{document}

\title{Reliable Interval Prediction of Minimum Operating Voltage Based on On-chip Monitors via Conformalized Quantile Regression
}

\author{
\IEEEauthorblockN{Yuxuan Yin}
\IEEEauthorblockA{\textit{Dept.  of ECE} \\
\textit{University of California}\\
Santa Barbara, USA \\
y\_yin@ucsb.edu}
\and
\IEEEauthorblockN{Xiaoxiao Wang}
\IEEEauthorblockA{\textit{NXP Semiconductors} \\
Austin, USA  \\
Xiaoxia3@andrew.cmu.edu
}
\and
\IEEEauthorblockN{Rebecca Chen}
\IEEEauthorblockA{\textit{NXP Semiconductors} \\
Austin, USA \\
rebecca.chen\_1@nxp.com}
\and
\IEEEauthorblockN{Chen He}
\IEEEauthorblockA{\textit{NXP Semiconductors} \\
Austin, USA \\
chen.he@nxp.com}
\and
\IEEEauthorblockN{Peng Li}
\IEEEauthorblockA{\textit{Dept.  of ECE} \\
\textit{University of California}\\
Santa Barbara, USA \\
lip@ucsb.edu}
}

\maketitle

\begin{abstract}
Predicting the minimum operating voltage ($V_{min}$) of chips is one of the important techniques for improving the manufacturing testing flow, as well as ensuring the long-term reliability and safety of in-field systems. Current $V_{min}$ prediction methods often provide only point estimates, necessitating additional techniques for constructing prediction confidence intervals to cover uncertainties caused by different sources of variations. While some existing techniques offer region predictions, but they rely on certain distributional assumptions and/or provide no coverage guarantees. In response to these limitations, we propose a novel distribution-free $V_{min}$ interval estimation methodology possessing a theoretical guarantee of coverage. Our approach leverages conformalized quantile regression and on-chip monitors to generate reliable prediction intervals. We demonstrate the effectiveness of the proposed method on an industrial 5nm automotive chip dataset. Moreover, we show that the use of on-chip monitors can reduce the interval length significantly for $V_{min}$ prediction.
\end{abstract}

\begin{IEEEkeywords}
chip performance prediction, on-chip monitors, conformal prediction, quantile regression
\end{IEEEkeywords}

\section{Introduction}
Measurement of the minimum operating voltage ($V_{min}$) is one of the important testing procedures to determine chip performance. It facilitates the detection of inferior products, the conservation of power consumption, and the indication of potential early life failures. 
As technology nodes keep scaling, $\Vmin$ tests via structural test patterns (e.g., SCAN) become more and more crucial and necessary to screen out tiny flaws and defects \cite{VminTest} inside chips. 

Conventional $\Vmin$ measurements involve testing chips at a high operating voltage and decreasing step by step until they fail, which is time-consuming. Moreover, such a strategy is exclusively applicable in the manufacturing test process, but not in-field systems. To this end, researchers propose to build machine learning based $\Vmin$ predictors utilizing low-cost features, such as parametric testing data from the production test flow and on-chip monitor data for the in-field prediction \cite{RODesign, StructualFmax, ML-Assisted, yin-mln-itc}. Many regression models have been explored recently, including linear regression \cite{ML-Assisted}, Gaussian Process (GP) \cite{StructualFmax}, and Neural Network (NN) \cite{yin-mln-itc}.
For instance, Chen demonstrated a low-cost approach to predict the system $\Fmax$ (the maximum operating frequency) using the structural $\Fmax$ of flip flops \cite{StructualFmax} via a GP model, whose kernel hyperparameter length scales are used as indicators of the significance of features. Yin adopted a constrained NN to capture the monotonicity between RO delay and $\Vmin$ degradation \cite{yin-mln-itc}.
Although these methods provide promising point estimation for $\Vmin$, additional techniques are still required to construct prediction intervals to ensure high coverage of true $\Vmin$ to account for the uncertainties due to variations of process, voltage, temperature, operating frequency, application mode, etc.

Uncertainty Quantification (UQ) for machine learning provides the model's confidence interval. 
Commonly employed UQ methods include 1) Bayesian approaches such as GP \cite{GP} and Bayesian neural networks \cite{BNN}, 2) neural networks ensemble \cite{Ensemble}, and 3) Quantile Regression (QR) \cite{QR}.
While these methods excel at estimating uncertainty within the training data distribution, their prediction intervals often lack a reliable coverage guarantee for new testing data. Consequently, none of these approaches fully meet the stringent demands of the silicon industry for generating robust $\Vmin$ intervals to ensure high reliability.

\begin{table}[!tbp]
    \centering
    \caption{Comparison of uncertainty quantification methods}
    \begin{tabular}{l c c c c c }
    \toprule
       Property           &  Bayesian & Ensemble & QR    & CP          & CQR       \\
   \midrule
       Distribution-free            & \xmark    & \cmark  & \cmark & \cmark  & \cmark  \\ 
       Agnostic model               & \xmark    & \xmark  & \cmark & \cmark  & \cmark  \\ 
       \makecell[l]{Coverage guarantee \\ for test data}           & \xmark    & \xmark  & \xmark & \cmark  & \cmark  \\ 
       \makecell[l]{Adaption to \\ heteroscedasticity}           & \cmark    & \cmark  & \cmark & \xmark  & \cmark  \\ 
       Computational efficiency     & \xmark    & \xmark  & \cmark & \cmark  & \cmark  \\ 
\bottomrule
    \end{tabular}
    \label{tab:uq-methods-properties}
\end{table}

Conformal Prediction (CP) \cite{CP} emerges as a promising distribution-free UQ method for constructing intervals based on any point predictor while offering a nonasymptotic coverage guarantee. CP leverages a calibration dataset to assess the uncertainty associated with a fitted regression model by analyzing its prediction residuals. However, vanilla CP exhibits limitations as a $\Vmin$ region predictor, as it constructs constant intervals for all testing samples, potentially leading to excessive margins for normal chips and inadequate coverage for anomalous ones.

To this end, we propose a distribution-free $\Vmin$ interval prediction framework with a theoretical coverage guarantee. Our approach leverages 
Conformalized Quantile Regression (CQR) and on-chip monitors to construct prediction intervals. 
Our primary contributions are outlined as follows:

    $\smallbullet$ We conduct a comprehensive comparison among various $\Vmin$ point predictors for our industrial dataset. 
    We discover that while no golden model outperforms others for all scenarios, the prediction accuracy of linear regression is competitive overall.
    Moreover, on-chip monitors are capable of predicting future $\Vmin$ degradation.
    
    $\smallbullet$ We introduce CQR to the context of $\Vmin$ interval estimation, showcasing its better performance in terms of coverage rate and interval length when compared to alternative UQ models.

    $\smallbullet$ Through empirical analysis, we demonstrate that the inclusion of on-chip monitor data yields substantial improvements in the precision of interval predictions. 




\begin{figure}[tb]
    \centering
    \includegraphics[width=0.48\textwidth]{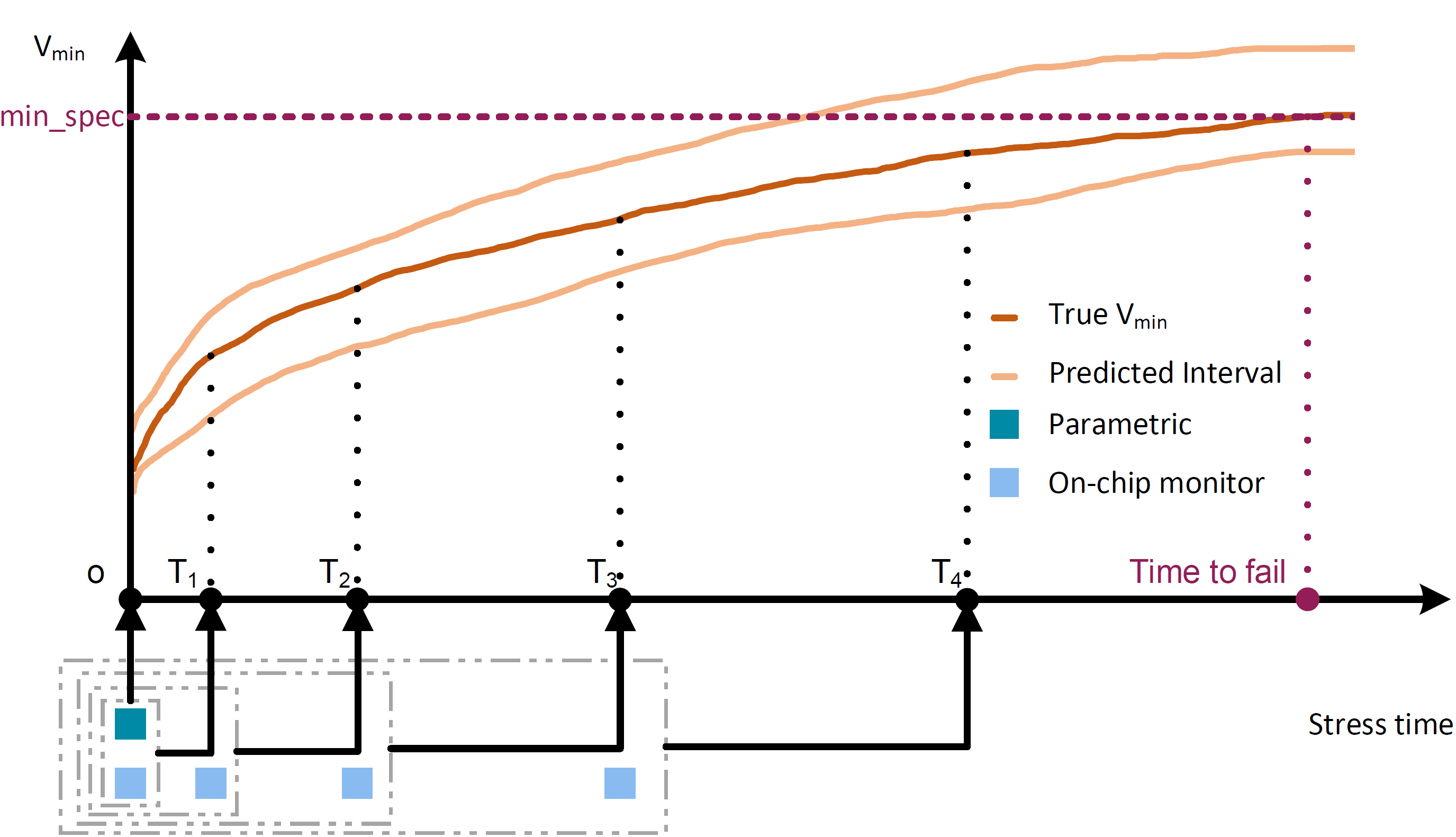}
    \caption{$\Vmin$ prediction flow}
    \label{fig:diagram}
\end{figure}

\section{Preliminaries}

\subsection{Point Prediction}
For the task of $\Vmin$ point estimation, in both the product flow and in-field scenarios, the objective remains consistent: utilizing a set of features to predict a single value. We denote these features as a $D$ dimension vector $\x \in \RD$, the $\Vmin$ as a real number $\y \in \R$, and the point predictor as $\modelPoint(\cdot;\paramModel):\RD \to \R$, parameterized by $\paramModel$. Given a training dataset of $N$ tested chips $\traindataset=\{(\x_i, \y_i)\}_{i=1}^N$, the predictor is optimized by minimizing the mean of a loss function $\lossp$:
\begin{equation} \label{eq:obj}
    \paramModel^* = \argmin_{\paramModel} \lossp \big( \modelPoint(\X;\paramModel), \Y \big),
\end{equation}
where $\X = [ \x_1, \cdots, \x_N]^T \in \R^{N \times D}$ is a matrix of inputs, and $\Y = [ \y_1, \cdots, \y_N]^T \in \R^{N}$ is a vector of true $\Vmin$.

\subsection{Region Prediction}


In manufacturing test processes, engineers often face risks of over-kill or under-kill when relying solely on $\Vmin$ point predictions to identify abnormal products due to process variations. In in-field scenarios, point estimation can be highly unreliable due to the presence of numerous environmental uncertainties. Consequently, the utilization of prediction intervals becomes essential for effectively detecting outliers and identifying potential failures. 

Unlike point estimation, which only generates a single value for an input example, region prediction provides an interval prediction. A region regressor $\modelRegion(\cdot;\paramLow, \paramHigh):\RD \to \R^2$, consisting of a pair $\cilow (\cdot; \paramLow):\RD \to \R$ and $\cihigh (\cdot;\paramHigh):\RD \to \R$ of the lower and the upper bound function, maps a sample $\x$ to a closed region $C(\x)$:
\begin{equation}
    C(\x) = \big[\cilow (\x; \paramLow), \quad \cihigh (\x; \paramHigh) \big].
\end{equation}
Given a coverage rate $1-\coverage$ where $\coverage \in [0,1]$ and the training dataset $\traindataset$, the prediction intervals of a region regressor $\modelRegion$ should be able to cover at least $1-\coverage$ labels:
\begin{equation} \label{eq:coverage}
    \Prob \big\{ \y \in C(\x) | (\x, \y) \in \traindataset \big\} \geq 1 - \coverage.
\end{equation}

We introduce two well-known region regression methods satisfying \cref{eq:coverage}: Gaussian process and quantile regression. Their theoretical traits are summarized in \cref{tab:uq-methods-properties}. 

\subsubsection{Gaussian Process (GP)} GP is a non-parametric Bayesian method that provides a posterior Gaussian distribution for any testing point \cite{GP}. Suppose the posterior mean is $\mu (\x) \in \R$ and the posterior variance is $\sigma^2 (\x) \geq 0$ for sample $\x$, we are able to construct an interval $C(\x)$ satisfying \cref{eq:coverage}:
\begin{equation}\label{eq:ci-gp}
    C(\x) = \big[\mu (\x) + K_{lo}\sigma (\x), \quad \mu (\x) + K_{hi}\sigma (\x) \big],
\end{equation}
where $K_{lo} = \Phi^{-1}(\coverage/2) < 0 $, $K_{hi} = \Phi^{-1}(1-\coverage/2) > 0 $, and $\Phi$ is the cumulative distribution function of the standard Gaussian distribution. 

\subsubsection{Quantile Regression (QR)} Apart from traditional regression analysis with Mean Square Error (MSE) loss that estimates the conditional mean of $\Vmin$, QR estimates the conditional quantile \cite{QR}. Given a quantile $q \in [0,1]$, a QR model is trained to minimize the quantile loss \cite{QR} $\lossq$ in \cref{eq:obj}:
\begin{equation}\label{eq:ci-qr}
    \lossq \big( \y, \hat{\y} \big) := \max \big\{ q( \y - \hat{\y} ), (1-q)(  \hat{\y} -\y )  \big\},
\end{equation}
where $\hat{\y} = \modelPoint(\x;\paramModel)$ is the prediction of quantile $\Vmin$.

By selecting two different quantiles $\qlow = \coverage/2$ and $\qhigh = 1- \coverage/2$, we can train two quantile regressors, the interval between which achieves the coverage in \cref{eq:coverage}.

QR can be easily added to any point regressor where its objective is to minimize the MSE loss by applying the pinball loss instead. 

\section{Methodology}
\subsection{Overview of $\Vmin$ Prediction}
Our $\Vmin$ prediction framework is depicted in \cref{fig:diagram}, where four stress read points are drawn for illustration. $\Vmin$ at each stress read point will be predicted. The horizontal dash line (min\_spec) stands for the product specification of the minimum operating voltage, i.e., device with $\Vmin$ higher than that threshold will violate the specification and likely become a failure. 

We utilize low-cost parametric data and on-chip data to predict $\Vmin$ at time zero and subsequent read points during stress simulated in-field life. Note that stress is done at an elevated voltage such that a much shorter stress duration is equivalent to a much longer in-field life. Specifically, two kinds of $\Vmin$ prediction scenarios are considered: in the production test flow, and in the in-field deployment which is simulated by accelerated stress. In the first case, both production parametric test data and on-chip data are included to build $\Vmin$ predictors. In the second case, however, we make $\Vmin$ degradation prediction based on all accessible features before the $\Vmin$ test timestamp, including production parametric test data at time zero and on-chip data measured at all previous read points during stress. In our industrial dataset, both $\Vmin$ and on-chip data are collected at the same read point, and the total number of read points is relatively small, i.e., less than 10. In this case, time series methods would suffer over-fitting problems. Thus, we treat on-chip data at different read points as different features, and apply CQR to predict $\Vmin$ intervals.

Since CQR is originated from CP, we first briefly summarize how CP works, and then present CQR for $\Vmin$ interval prediction.

\subsection{Conformal Prediction (CP)}
Even though the coverage of prediction intervals is guaranteed for the training dataset $\traindataset$ in GP and QR, such characteristic is not held for a testing instance $(\x_{N+1}, \y_{N+1})$: 
\begin{equation} \label{eq:conformal-coverage}
    \Prob \big\{ \y_{N+1} \in C(\x_{N+1}) \big\} \geq 1 - \coverage.
\end{equation}
The adoption of the aforementioned two region predictors for new examples is risky without the coverage guarantee. 

In semiconductor industry, all chips can be viewed as examples from a hidden distribution: $\{(\x_i, \y_i)\}_{i=1}^{N+1}$ are sampled i.i.d. from a distribution $P_{XY}$. CP can help to calibrate any heuristic interval to meet the coverage guarantee in \cref{eq:conformal-coverage} \cite{CP}. CP has two main versions: \textit{full} CP and \textit{split} CP. In regression tasks, full CP needs infinite times of model fitting, rendering it impossible for practical usage. On the contrary, split CP is more computationally efficient with the scarification of splitting the training dataset.

We outline how split CP utilizes a $\Vmin$ point predictor $\modelPoint$ to generate a interval $C(\x)$ for $\x$:

$\smallbullet$ Split the training dataset $\traindataset$ into a new training dataset $\traindataset_{tr}$, and a small calibration dataset $\traindataset_{ca}$ such that $\traindataset_{tr} \cup \traindataset_{ca} = \traindataset$, and $\traindataset_{tr} \cap \traindataset_{ca} = \phi$.

$\smallbullet$ Fit the point regressor $\modelPoint$ in $\traindataset_{tr}$.

$\smallbullet$ Compute $\hat{q}$ as the $\lceil (M+1)(1-\coverage) \rceil /M\text{-th}$ quantile of the conformal score function $s(\x, \y)$ of absolute residuals in the calibration set $\traindataset_{ca}$:
\begin{equation}
    s(\x, \y) = |\y - \modelPoint(\x;\paramModel)|,
\end{equation}
where $M$ is the number of examples in $\traindataset_{ca}$.

$\smallbullet$ Construct the interval for $\x_{N+1}$, satisfying \cref{eq:conformal-coverage}:
\begin{equation}\label{eq:ci-cqr}
C(\x_{N+1}) = \big[\modelPoint(\x_{N+1};\paramModel) -\hat{q}, \quad \modelPoint(\x_{N+1};\paramModel) +\hat{q} \big].
\end{equation}

\subsection{Conformalized Quantile Regression (CQR)}
While split CP satisfies the coverage guarantee, the length of predicted intervals is $2\hat{q}$, remaining fixed to different inputs. This property may incur overkill for good products and underkill for defective ones.
CQR, however, is a variant interval prediction method combining CP and QR together. 

We describe the procedures of \textit{split} CQR:

$\smallbullet$ Split the training dataset $\traindataset$ .

$\smallbullet$ Fit the quantile regressor $\modelRegion$ in $\traindataset_{tr}$.

$\smallbullet$ Compute $\hat{q}$ as the $\lceil (M+1)(1-\coverage) \rceil /M\text{-th}$ quantile of the conformal score function $s(\x,\y)$ in $\traindataset_{ca}$, where
\begin{equation}
    s(\x,\y) = \max \{ \cilow (\x; \paramLow) - \y, \quad  \y - \cihigh (\x; \paramHigh)\}.
\end{equation}

$\smallbullet$ Construct the interval for $\x_{N+1}$ satisfying \cref{eq:conformal-coverage}:
\begin{equation}
C(\x_{N+1}) = \big[\cilow (\x_{N+1}; \paramLow) -\hat{q}, \quad \cihigh (\x_{N+1}; \paramHigh) +\hat{q} \big].
\end{equation}

CQR inherits good features of CP and QR, as shown in \cref{tab:uq-methods-properties}. It is shown empirically effective in achieving the shortest interval length than CP and QR across 11 datasets while persisting the designed coverage rate \cite{CQR}. Herein, we adopt it for reliable $\Vmin$ interval prediction.

\section{Experimental Results}
\subsection{Industrial Dataset}
\begin{table}[tbp]
    \centering
    \caption{Input feature description}
    \begin{tabular}{l c c c}
    \toprule
       Attribute  &  Parametric & On-chip (ROD) & On-chip (CPD) \\
   \midrule
       Quantity & 1800 & 168 & 10 \\
     Temperature (\degree C) & -45, 25, 125 & 25 & 80 \\
     Read point (hour) & 0 & \multicolumn{2}{c}{0, 24, 48, 168, 504, 1008} \\
    \bottomrule
          
    \end{tabular}
    \label{tab:dataset}
\end{table}
Our experiments use 156 5nm automotive chips to demonstrate the effectiveness of the proposed $\Vmin$ prediction framework.  As shown in \cref{fig:diagram}, parametric data and on-chip monitor data are considered for $\Vmin$ prediction. We describe how the input features and the output $\Vmin$ are collected.

All 156 chips go through the dynamic Dhrystone stress at elevated voltage in Burn-In (BI) oven for 1008 hours to simulate in-field long-term aging degradation. At specific stress read points, i.e., 0, 24, 48, 168, 504, and 1008 hours, we pause the stress process and 1) test SCAN $\Vmin$, 2) perform the parametric tests, and 3) collect on-chip monitor data. SCAN $\Vmin$ is tested on Automatic Test Equipment (ATE) tester, at temperatures of -45\degree C, 25\degree C, and 125\degree C. The parametric tests are also performed on ATE tester, including IDDQ, trip IDD, leakage, etc., across all three temperatures. The chip has two types of on-chip monitors: domain sensors which include Ring Oscillator Delay (ROD) sensors and in-situ Critical Path Delay (CPD) sensors. In our experiment, due to hardware and logistic process limitations, ROD is measured on ATE at room temperature (25\degree C) only while CPD is measured in-situ in BI oven at 80C. We summarize the traits of input features in \cref{tab:dataset}.

\begin{figure}[htb]
    \centering
    \includegraphics[width=0.48\textwidth]{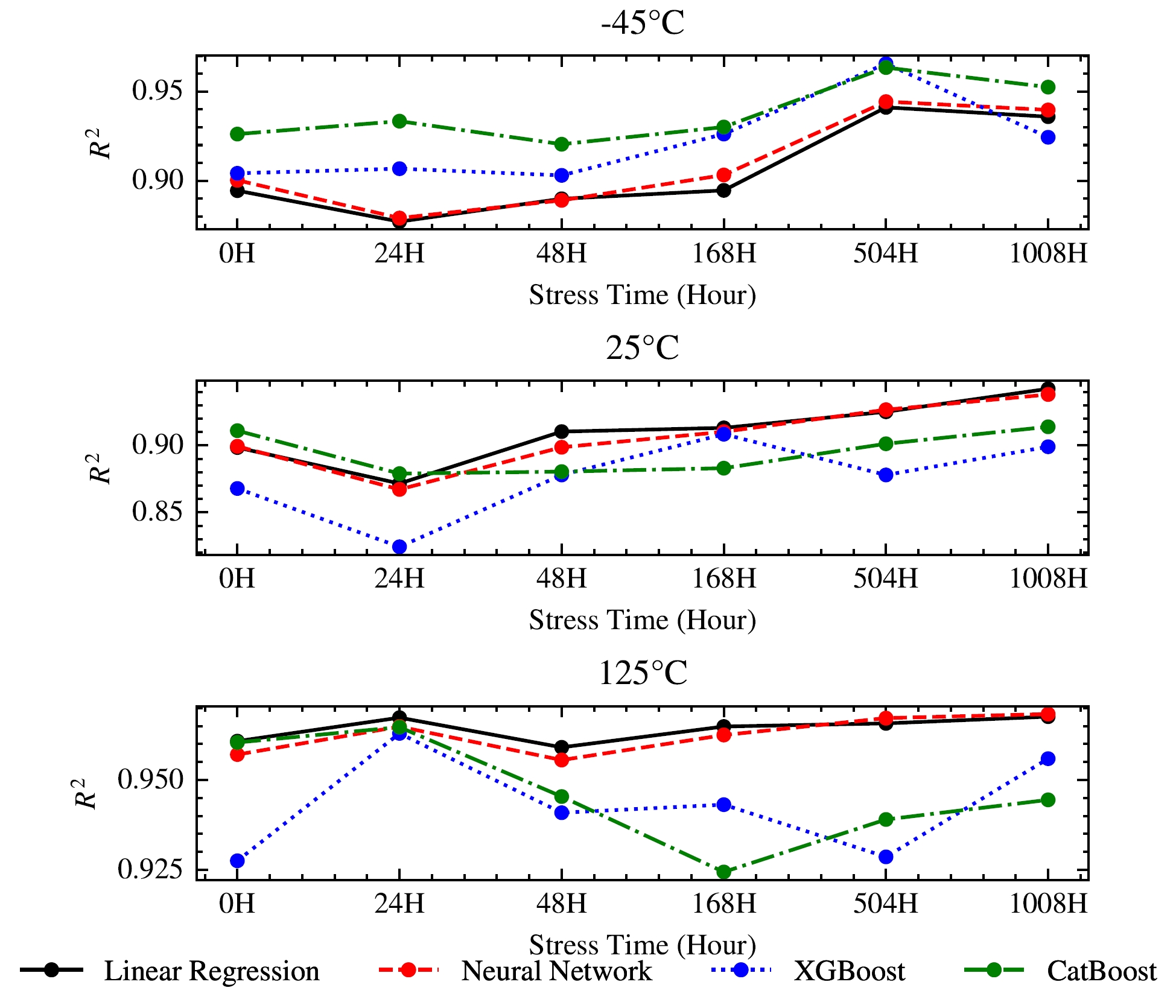}
    \caption{SCAN $\Vmin$ point prediction}
    \label{fig:dcscan-point}
\end{figure}

\subsection{Experimental Settings}
We illustrate the features used for $\Vmin$ prediction at each read point and the evaluation metrics for point prediction and interval regression.
As shown in \cref{fig:diagram}, for the prediction of $\Vmin$ at time 0, both parametric test data and on-chip monitor data collected at time 0 are utilized to predict $\Vmin$; For the prediction of $\Vmin$ at the subsequent read points to enable in-field failure prediction, we use on-chip monitor data collected at all previous read points and parametric data collected at time 0, because parametric tests are no longer possible once chips are shipped to customers and deployed in-field.

For $\Vmin$ point prediction, the performance criteria are the coefficient of determination ($\Rtwo$) and Root Mean Square Error (RMSE); For $\Vmin$ region prediction, the metrics are the average interval length and the coverage of true $\Vmin$ of the testing data.

To reduce the influence of randomization, a 4-fold cross-validation is adopted. We report the average score of each metric across the 4 testing folds. In CQR, 75\% training data are used to train predictors while the remaining 25\% chips are held for calibration. To ensure a fair comparison, we use the same random seed for all $\Vmin$ interval predictors.

\subsection{Descriptions of $\Vmin$ Point Regressors} \label{sec:config-point}
ML models with fewer learnable parameters and simpler structures are more favorable for our high-dimensional small data scenario. Moreover, feature selection is an essential dimension reduction technique for some ML models to avoid overfitting problems.

Firstly, we demonstrate model selection for $\Vmin$ point prediction. 5 regressors are considered: Linear Regression (LR), Gaussian Process (GP) \cite{GP}, XGBoost \cite{XGBoost}, CatBoost \cite{CatBoost}, and a 2-layer Neural Network (NN). The detailed configurations of each regressor except LR are provided below:

\subsubsection{Gaussian Process} GP utilizes a radial basis function kernel, whose parameters are optimized to maximize the likelihood of training data.

\subsubsection{XGBoost} We utilize the default hyper-parameters in the XGBoost Python package. 

\subsubsection{CatBoost} We utilize the default hyperparameters in the CatBoost Python package except for one hyper-parameter: the number of boosting trees. The default number is 1000, which seems too large for our small dataset including 156 chips, and potentially causes over-fitting.  Therefore, we reduce it to 100.

\subsubsection{Neural Network} 
We consider a shallow fully-connected multilayer perceptron (MLP) with one hidden layer containing 16 neurons with Rectified Linear Units (ReLU) \cite{ReLU} activation functions. The optimizer is Adam \cite{kingma2014adam} whose learning rate is 0.01, the number of epochs is 3000, and the weight of $L_2$ penalty is 0.1. These configurations are the same as \cite{yin-mln-itc}.

Then, we discuss how to select a small set of informative features among thousands of input data. For XGBoost and CatBoost which have an intrinsic feature selection mechanism, all raw data are directly fed to regressors. For the rest of the three methods, we apply Correlation Feature Selection (CFS) \cite{CFS} with the Pearson correlation 
to pick 1 to 10 features as input data and report the best testing scores.

\subsection{$\Vmin$ Point Prediction Results} \label{sec:result-point}
The $R^2$ of $\Vmin$ point predictions of regression models are depicted in \cref{fig:dcscan-point} 
For SCAN $\Vmin$ tested at time 0, while CatBoost is the best method across all three temperatures, linear regression is also performing well with a small drop of $R^2$, which is less than 0.03. 
For all methods except GP, the RMSE for $\Vmin$ point predictions are within $2.5mV$ to $7mV$ ($12mV$ to $22mV$ for GP) for all scenarios, and exhibiting similar comparison as $R^2$ among different models, i.e.,  CatBoost performs best for time 0 prediction while linear regression performs reasonably well overall. As linear regression is straightforward to implement by either software or hardware, it is a sufficiently good option for $\Vmin$ time 0 prediction in industrial production tests.

For $\Vmin$ degradation prediction, no regression model is outperforming the rest across all temperatures and stress read points, in terms of $R^2$ and RMSE. We note that linear regression is still performing reasonably well, and even the best one for predicting SCAN $\Vmin$ at 25\degree C and 125\degree C, for both $R^2$ and RMSE. With its simplicity, implementing a linear regression model with an on-chip hardware accelerator seems to be a viable option for in-field $\Vmin$ degradation prediction.

In addition, an interesting observation is that there is no clear reduction of $R^2$ in SCAN $\Vmin$ degradation prediction accuracy from 0 to 1008 hours. It demonstrates that our design of on-chip monitors captures informative gate-level features that exhibit a strong correlation with system-level $\Vmin$. 

\begin{table*}[!bp]
    \centering
    \caption{Average length and coverage of prediction intervals for SCAN $\Vmin$ across 156 chips}
    \begin{tabular}{c l c c c c c c }
    \toprule
       \multirow{2}{*}{\makecell{Stress Time  \\ (Hour)}}           &  \multirow{2}{*}{Method} & \multicolumn{2}{c}{-45\degree C}& \multicolumn{2}{c}{25\degree C}& \multicolumn{2}{c}{125\degree C} \\
       && Length ($mV$) & Coverage (\%)  & Length ($mV$) &  Coverage (\%)  & Length ($mV$) & Coverage (\%)    \\
   \midrule
       \multirow{9}{*}{0}    & GP & 61.96 & 85.9 & 48.56 & 93.59 & 51.88 & 89.1    \\
 & QR Linear Regression & 51.0 & 91.03 & 14.14 & 83.33 & 15.98 & 83.33    \\
 & QR Neural Network & 30.44 & 66.84 & 18.28 & 53.91 & 21.33 & 52.83    \\
 & QR XGBoost & 50.31 & 51.28 & 28.22 & 89.1 & 30.96 & 82.05    \\
 & QR CatBoost & 2.48 & 10.26 & 0.98 & 14.1 & 1.37 & 24.36    \\
 & CQR Linear Regression & 53.76 & 92.95 & 17.37 & 95.51 & 19.39 & 91.03    \\
 & CQR Neural Network & 114.3 & 94.81 & 52.75 & 93.11 & 77.54 & 94.01    \\
 & CQR XGBoost & 60.84 & 95.51 & 31.91 & 92.95 & 48.48 & 98.72    \\
 & \textbf{CQR CatBoost} & \textbf{24.11} & \textbf{91.67} &\textbf{13.94} & \textbf{92.95} & \textbf{12.72} & \textbf{91.67}    \\
   \midrule
       \multirow{9}{*}{24}    & GP & 56.76 & 84.93 & 48.64 & 94.87 & 50.53 & 87.74    \\
 & QR Linear Regression & 26.7 & 85.62 & 18.3 & 80.13 & 13.28 & 85.16    \\
 & QR Neural Network & 24.19 & 68.67 & 16.33 & 49.52 & 19.78 & 53.68    \\
 & QR XGBoost & 43.27 & 39.04 & 32.64 & 87.18 & 30.28 & 86.45    \\
 & QR CatBoost & 1.54 & 3.42 & 1.38 & 19.87 & 1.77 & 20.65    \\
 & CQR Linear Regression & 43.1 & 99.32 & 20.68 & 89.74 & 17.07 & 95.48    \\
 & CQR Neural Network & 117.82 & 97.01 & 53.66 & 93.34 & 84.99 & 95.45    \\
 & CQR XGBoost & 65.3 & 99.32 & 43.5 & 92.95 & 42.41 & 92.9    \\
 & \textbf{CQR CatBoost} & \textbf{27.1} &\textbf{97.95} & \textbf{16.58} & \textbf{94.87} &\textbf{15.34} & \textbf{93.55}    \\
   \midrule
       \multirow{9}{*}{48}    & GP & 56.83 & 81.13 & 49.84 & 89.72 & 53.84 & 82.24    \\
 & QR Linear Regression & 29.77 & 84.91 & 20.03 & 81.31 & 13.98 & 82.24    \\
 & QR Neural Network & 29.66 & 68.04 & 44.71 & 92.05 & 26.14 & 50.79    \\
 & QR XGBoost & 45.43 & 45.28 & 35.78 & 85.98 & 48.6 & 84.11    \\
 & QR CatBoost & 1.64 & 11.32 & 1.07 & 16.82 & 1.79 & 19.63    \\
 & CQR Linear Regression & 36.92 & 93.4 & \textbf{29.34} & \textbf{94.39} & \textbf{20.61} & \textbf{93.46}    \\
 & CQR Neural Network & 100.62 & 95.59 & 58.75 & 95.62 & 80.64 & 95.07    \\
 & CQR XGBoost & 62.81 & 98.11 & 49.82 & 94.39 & 55.12 & 95.33    \\
 & \textbf{CQR CatBoost }& \textbf{24.3} &\textbf{95.28} & \textbf{29.61} &\textbf{96.26} & \textbf{19.23}& \textbf{89.72}    \\
   \midrule
       \multirow{9}{*}{168}    & GP & 54.45 & 79.81 & 50.43 & 84.91 & 54.42 & 85.58    \\
 & QR Linear Regression & 26.05 & 81.73 & 44.0 & 89.62 & 12.27 & 81.73    \\
 & QR Neural Network & 27.74 & 72.68 & 43.56 & 84.12 & 26.03 & 48.32    \\
 & QR XGBoost & 38.27 & 75.96 & 39.89 & 84.91 & 49.65 & 85.58    \\
 & QR CatBoost & 1.81 & 19.23 & 0.71 & 13.21 & 1.78 & 20.19    \\
 & CQR Linear Regression & 36.28 & 92.31 & 51.35 & 94.34 & \textbf{17.09} & \textbf{89.42}    \\
 & CQR Neural Network & 82.98 & 95.33 & 60.16 & 95.48 & 80.99 & 95.42    \\
 & CQR XGBoost & 56.65 & 96.15 & 48.61 & 94.34 & 57.75 & 92.31    \\
 & \textbf{CQR CatBoost} & \textbf{28.71} &\textbf{93.27} & \textbf{20.49} & \textbf{91.51} & \textbf{20.49} & \textbf{92.31}    \\
   \midrule
       \multirow{9}{*}{504}    & GP & 52.61 & 77.0 & 52.63 & 88.46 & 54.23 & 79.61    \\
 & QR Linear Regression & 25.46 & 83.0 & 37.71 & 88.46 & 26.14 & 88.35    \\
 & QR Neural Network & 25.51 & 70.39 & 46.33 & 92.16 & 48.65 & 83.49    \\
 & QR XGBoost & 35.9 & 78.0 & 43.14 & 84.62 & 47.71 & 83.5    \\
 & QR CatBoost & 1.43 & 12.0 & 1.54 & 18.27 & 2.24 & 20.39    \\
 & CQR Linear Regression & 31.2 & 91.0 & 45.21 & 93.27 & 32.05 & 94.17    \\
 & CQR Neural Network & 66.13 & 93.37 & 53.44 & 92.79 & 72.25 & 94.76    \\
 & CQR XGBoost & 46.81 & 93.0 & 46.83 & 87.5 & 58.74 & 96.12    \\
 &\textbf{CQR CatBoost} &\textbf{21.17} & \textbf{96.0} &\textbf{19.01} &\textbf{92.31} & \textbf{16.15} &\textbf{94.17}    \\
   \midrule
       \multirow{9}{*}{1008}    & GP & 53.18 & 78.12 & 52.45 & 91.84 & 53.22 & 82.65    \\
 & QR Linear Regression & 29.75 & 88.54 & 42.63 & 88.78 & 32.28 & 80.61    \\
 & QR Neural Network & 20.2 & 50.3 & 19.89 & 39.14 & 31.47 & 51.9    \\
 & QR XGBoost & 37.18 & 79.17 & 45.19 & 84.69 & 46.0 & 82.65    \\
 & QR CatBoost & 1.72 & 17.71 & 1.64 & 13.27 & 1.89 & 24.49    \\
 & CQR Linear Regression & 32.3 & 89.58 & 47.25 & 94.9 & 36.53 & 91.84    \\
 & CQR Neural Network & 78.55 & 98.2 & 66.8 & 93.08 & 65.86 & 92.25    \\
 & CQR XGBoost & 44.14 & 89.58 & 47.11 & 91.84 & 51.44 & 96.94    \\
 & \textbf{CQR CatBoost} &\textbf{17.64} & \textbf{93.75} & \textbf{18.7} &\textbf{94.9} & \textbf{14.68} &\textbf{89.8}    \\
\bottomrule
    \end{tabular}
    \label{tab:ci-prediction}
\end{table*}

\subsection{Descriptions of $\Vmin$ Region Regressors}
We consider three interval prediction methods: GP, QR, and CQR. QR and CQR are built on 4 point regressors: LR, NN, XGBoost, and CatBoost. The configurations of these models are the same as those in \cref{sec:config-point}.
We set $\coverage=0.1$ and let predictors generate an interval with 5\% to 95\% coverage.

\subsection{$\Vmin$ Region Prediction Results} \label{sec:result-region}

\begin{figure}[htb]
    \centering
    \includegraphics[width=0.48\textwidth]{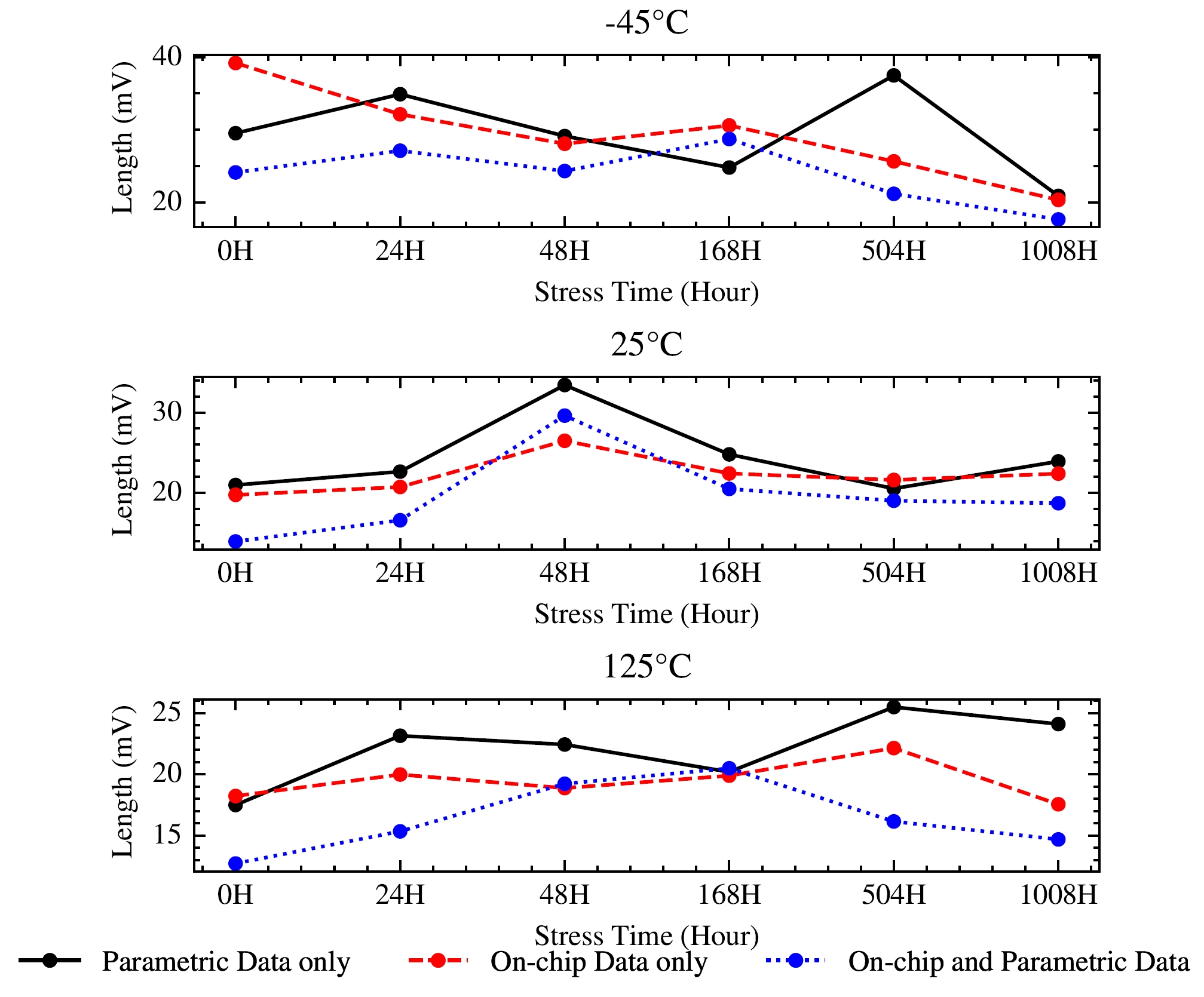}
    \caption{The average interval length of CQR CatBoost for SCAN $\Vmin$ prediction}
    \label{fig:dcscan-region}
\end{figure}

The average length of prediction intervals of SCAN $\Vmin$  and coverage rates are shown in \cref{tab:ci-prediction}. Both GP and QR underestimate the interval for testing chips, failing to meet the designed coverage rate. CQR, in contrast, successfully calibrates the undercovered interval predictions of QR across all stress read points and temperatures, underscoring the importance of applying conformal prediction for reliable region predictions.

CQR performs differently with different point regression models. The best variant is CQR CatBoost, achieving the shortest intervals with around 90\% coverage rate. While LR is competitive for point prediction in \cref{sec:result-point}, its CQR version predicts larger intervals than CQR CatBoost, especially for SCAN $\Vmin$ at -45\degree C and 25\degree C.

\begin{table}[!tbp]
    \centering
    \caption{ SCAN $\Vmin$ interval prediction via CQR CatBoost averaged across all stress time read points}
    \begin{tabular}{l c c c c}
    \toprule
       \multirow{2}{*}{Feature type} & \multicolumn{4}{c}{Avg Interval Length ($mV$)} \\
       &  -45\degree C & 25\degree C & 125\degree C  & Average  \\
   \midrule
       Parametric  &  29.44 & 24.38  & 22.14 & 25.32\\
        On-chip  &  29.32   & 22.22 & 19.44  & 23.66\\
        On-chip and Parametric  &  \textbf{23.84} & \textbf{19.72} & \textbf{16.43} & \textbf{20.00}\\
    \midrule
        On-chip monitor gain &  19.02\%   & 19.11\% &  25.79\% & 21.01\% \\
    \bottomrule
    \end{tabular}
    \label{tab:on-chip-benefit}
\end{table}

\subsection{Benefits of On-chip Monitors}
We present evidence supporting the value of on-chip monitor data in the prediction of $\Vmin$ intervals. \cref{fig:dcscan-region} illustrates the interval length of CQR CatBoost with three types of feature sets: 1) parametric test data and on-chip monitor data (same to \cref{sec:result-region}), 2) parametric test data only, and 3) on-chip monitor data only. In addition, \cref{tab:on-chip-benefit} summarizes the average length across all read points of SCAN $\Vmin$ during stress.

Compared to utilizing parametric data only, the inclusion of on-chip monitor data results in a reduction of 21.01\% in the average interval length. 
Intriguingly, a CQR CatBoost model relying solely on on-chip monitor data outperforms the same model using only parametric test data, despite the much larger number of parametric data (\cref{tab:dataset}). This implies the on-chip monitor data could contain more information that facilitates $\Vmin$ estimation.

\section{Conclusion}

We propose a distribution-free $\Vmin$ interval estimation framework possessing a statistical coverage guarantee. 
By harnessing CQR in conjunction with on-chip monitor data, our approach achieves an average interval length of $20mV$ with a 90\% coverage rate for true $\Vmin$ values on our industrial dataset.
In the future, we will explore how to embed the proposed method 1) in the production test flow to accelerate the $\Vmin$ test and enhance the yield while screening out outliers, and 2) in the in-field systems to secure long-term reliability and safety. 
\section*{Acknowledgment}
The content of this paper has been developed with the support of Grant No. 1956313 from the National Science Foundation (NSF) and has also received partial funding from a Long Term University (LTU) grant provided by NXP.
\bibliographystyle{IEEEtran}
\bibliography{IEEEabrv, ref}

\end{document}